\documentclass[prd,twocolumn,
showpacs,preprintnumbers,nofootinbib]{revtex4}
\usepackage[dvips]{graphicx}
\usepackage{enumerate}
\usepackage{amsmath,amssymb}
\usepackage{mathrsfs}
\usepackage[dvips]{graphicx}
\usepackage{bm}

\begin{document}
\preprint{CHIBA-EP-192, 2012}

\title{A unitary and renormalizable model for massive Yang-Mills fields 
\\
without Higgs fields}


\author{Kei-Ichi Kondo$^{1}$}

\affiliation{$^1$Department of Physics,  
Graduate School of Science, 
Chiba University, Chiba 263-8522, Japan
}
\begin{abstract}
We propose a massive Yang-Mills model blessed with both physical unitarity and renormalizability without Higgs particles. 
This  is achieved by a novel nonlinear but local transformation from the original fields in the Curci-Ferrari model to the massive vector field with the spin one, which has the correct physical degrees of freedom and invariant under an extended BRST transformation. 
We clarify the reason for failures of the preceding attempts and  check the physical unitarity in perturbation theory.

\end{abstract}

\pacs{12.38.Aw, 21.65.Qr}

\maketitle

Currently, the Higgs mechanism \cite{Higgs66} is widely accepted as a unique way for providing the masses for gauge bosons in quantum field theories, because it is the only one established method  which enables one to  maintain both renormalizability and unitarity \cite{tHooft71} in massive non-Abelian gauge field theories \cite{YM54}.   
Indeed, the Higgs particle becomes an indispensable ingredient in the unified model of electro-weak interactions, although it has not yet been  detected, but it is to be discovered in the near future.

There are continued attempts to construct an alternative massive non-Abelian gauge model \cite{DV70,SF70,Boulware70,CF76,CF76b,FMTY81,Ojima82,BSNW96,DTT88,RRA04}. 
Such a model is anxious to understand  mass gap and confinement caused by the strong interactions in QCD without the Higgs field  \cite{Cornwall82,decoupling}, apart from  the shortcomings in the Higgs approach, e.g., the indeterminate masses and couplings of the Higgs particles themselves.
However, all these efforts were unsuccessful in coping with both renormalizability and unitarity very well, see  \cite{DTT88,RRA04}  for reviews.

The aim of this paper is to provide a massive Yang-Mills model blessed with both unitarity and renormalizability without the Higgs fields, which is  written in polynomials of the fields (we exclude the nonpolynomial type \cite{FMTY81} from our discussions).
In fact, we start from the Curci-Ferrari (CF) model \cite{CF76}, which is a massive extension of the massless Yang-Mills theory in the most general renormalizable gauge having both the BRST and anti-BRST symmetries \cite{Baulieu85}.
In preceding studies for the CF model \cite{CF76,CF76b,Ojima82,BSNW96}, the CF model is proved to be renormalizable, whereas the CF model has been concluded to be nonunitary \cite{CF76b,Ojima82,BSNW96} when restricted to the physical subspace specified by the Kugo-Ojima (KO) subsidiary condition \cite{KO78}. 
In contrast to these results, we show that the CF model fulfills the physical unitarity, if the massive vector field is correctly identified. 
This is achieved by constructing a nonlinear but local transformation from the original fields to the physical massive vector field $\mathscr{K}_\mu$ which is invariant under the modified (anti)BRST transformation. 
We clarify the reason for failures of the preceding attempts from this viewpoint.  
We  demonstrate that the CF model rewritten in terms of the correct massive vector field satisfies the physical unitarity in a non-trivial lowest order of perturbation theory.

We consider the CF model with the total Lagrangian density written in terms of the Yang-Mills field $\mathscr{A}_\mu$, the Faddeev-Popov (FP) ghost field $\mathscr{C}$, antighost field $\bar{\mathscr{C}}$ and the Nakanishi-Lautrup (NL) field $\mathscr{N}$:
\begin{align}
 \mathcal{L}_{\rm mYM}^{\rm tot} &= \mathcal{L}_{\rm YM} + \mathcal{L}_{\rm GF+FP} + \mathcal{L}_{\rm m} ,
 \nonumber\\
 \mathcal{L}_{\rm YM} &= - \frac14 \mathscr{F}_{\mu\nu} \cdot \mathscr{F}^{\mu\nu} ,
 \nonumber\\
 \mathcal{L}_{\rm GF+FP} &= \mathscr{N} \cdot \partial^\mu \mathscr{A}_\mu  + \frac{\beta}{4}( \mathscr{N} \cdot \mathscr{N} + \bar{\mathscr{N}} \cdot \bar{\mathscr{N}} )
 \nonumber\\
  &+ i \bar{\mathscr{C}} \cdot \partial^\mu D_{\mu}[\mathscr{A}] \mathscr{C}  ,
\nonumber\\
 \mathcal{L}_{\rm m} &= \frac12 M^2 \mathscr{A}_\mu \cdot \mathscr{A}^\mu + \beta M^2 i \bar{\mathscr{C}} \cdot \mathscr{C} ,
 \label{mYM1}
\end{align}
where $\beta$ is a gauge-fixing parameter and $\bar{\mathscr{N}} := - \mathscr{N} + g i \bar{\mathscr{C}} \times \mathscr{C}$.
It is checked that $\mathcal{L}_{\rm YM} + \mathcal{L}_{\rm GF+FP}$ is invariant under both the usual BRST and anti-BRST transformations. This is not the case for the mass term $\mathcal{L}_{\rm m}$.  
Even in the presence of the mass term $\mathcal{L}_{\rm m}$, the  $\mathcal{L}_{\rm mYM}^{\rm tot}$ can be made invariant by modifying the BRST transformation \cite{CF76}:
$\delta_{\rm BRST}=\lambda {\boldsymbol \delta}$ with a Grassmannian number $\lambda$ and
\begin{align}
\begin{cases}
			{\boldsymbol \delta} \mathscr{A}_{\mu}(x) =  D_{\mu}[\mathscr{A}] \mathscr{C}(x) \\
			{\boldsymbol \delta} \mathscr{C}(x) 
=  -\frac{g}{2} \mathscr{C}(x) \times \mathscr{C}(x)  \\
			{\boldsymbol \delta} \bar{\mathscr{C}}(x) =  i \mathscr{N}(x) \\
			{\boldsymbol \delta} \mathscr{N}(x) =  M^2 \mathscr{C}(x)  
\end{cases} 
,
\end{align}
where 
$
 D_{\mu}[\mathscr{A}] \mathscr{C}(x) := \partial_{\mu} \mathscr{C}(x) - ig [\mathscr{A}(x), \mathscr{C}(x)] .
$
The modified BRST transformation reduces to the usual one in the limit $M \rightarrow 0$.
This is also the case for the modified anti-BRST transformation $\bar{\boldsymbol \delta}$ constructed similarly: 
\begin{align}
\begin{cases}
			\bar{\boldsymbol \delta} \mathscr{A}_{\mu}(x) = D_{\mu}[\mathscr{A}] \bar{\mathscr{C}}(x) \\
			\bar{\boldsymbol \delta} \bar{\mathscr{C}}(x) 
= -\frac{g}{2} \bar{\mathscr{C}}(x) \times \bar{\mathscr{C}}(x)  \\
			\bar{\boldsymbol \delta}  \mathscr{C}(x) = i \bar{\mathscr{N}}(x) \\
			\bar{\boldsymbol \delta} \bar{\mathscr{N}}(x) = - M^2 \bar{\mathscr{C}}(x)  
\end{cases} 
 .
\end{align}

We define the new vector field $\mathscr{K}_\mu$ written in terms of $\mathscr{A}_\mu$,  $\mathscr{C}$,  $\bar{\mathscr{C}}$ and $\mathscr{N}$:
\begin{align}
 \mathscr{K}_\mu :=&  \mathscr{A}_\mu - M^{-2} \partial_\mu \mathscr{N} 
- gM^{-2} \mathscr{A}_\mu \times \mathscr{N} 
\nonumber\\
&+ gM^{-2}  \partial_\mu \mathscr{C} \times i\bar{\mathscr{C}} 
+ g^2 M^{-2} (\mathscr{A}_\mu \times \mathscr{C}) \times i \bar{\mathscr{C}} .
\label{K}
\end{align}
In the Abelian limit, indeed, the field $\mathscr{K}$ reduces to the Proca field $U_\mu$,
i.e., 
$  \mathscr{K}_\mu  \rightarrow U_\mu := \mathscr{A}_\mu - M^{-2} \partial_\mu \mathscr{N}$.
In what follows, we show that the field $\mathscr{K}_\mu$ is identified with the non-Abelian version of the physical massive vector field with spin one, as assured by the properties:
\begin{enumerate}
\item[
(a)] $\mathscr{K}$ is invariant under the modified BRST transformation (off shell):
\begin{equation}
 {\boldsymbol \delta} \mathscr{K}_\mu(x) = 0 , 
\end{equation}
\item[
(b)] $\mathscr{K}$ is divergenceless (on shell):
\begin{equation}
 \partial^\mu \mathscr{K}_\mu(x) = 0 . 
\end{equation}
\end{enumerate}
The BRST invariance (a) is easily checked by using a more compact form:
\begin{align}
 \mathscr{K}_\mu =   \mathscr{A}_\mu + M^{-2} i {\boldsymbol \delta} \bar{\boldsymbol \delta} \mathscr{A}_\mu .
\label{K1}
\end{align}
Here it should be remarked that the modified (anti)BRST transformation is not nilpotent when  $M \not= 0$:
\begin{align}
\begin{cases}
{\boldsymbol \delta} {\boldsymbol \delta} \mathscr{A}_{\mu}(x) = 0 \\
{\boldsymbol \delta} {\boldsymbol \delta} \mathscr{C}(x) 
= 0  \\
{\boldsymbol \delta} {\boldsymbol \delta} \bar{\mathscr{C}}(x) = i M^2 \mathscr{C}(x) \\
{\boldsymbol \delta} {\boldsymbol \delta} \mathscr{N}(x) = - M^2 \frac{g}{2} \mathscr{C}(x) \times \mathscr{C}(x) \\
\end{cases} 
,
\end{align}
The divergenceless (b) is checked by using the field equation for $\mathscr{A}_\mu$:
\begin{equation}
 D^\nu[\mathscr{A}] \mathscr{F}_{\nu\mu} - \partial_\mu \mathscr{N} + g i\partial_\mu \bar{\mathscr{C}} \times \mathscr{C} + M^2 \mathscr{A}_\mu + gJ_\mu = 0 ,
\end{equation}
where
$
 J_\mu^A := g^{-1} \partial \mathscr{L}_{\rm matter}/\partial \mathscr{A}^{\mu A}
$.
This can be cast into the Abelian-like form:
\begin{equation}
 \partial^\nu  \mathscr{F}_{\nu\mu} + i {\boldsymbol \delta} \bar{\boldsymbol \delta} \mathscr{A}_\mu + M^2 \mathscr{A}_\mu + g\mathscr{J}_\mu = 0 ,
\end{equation}
thanks to introducing the color current $\mathscr{J}_\mu$  which is the conserved Noether current $\partial^\mu \mathscr{J}_\mu= 0$   associated with the color symmetry:  
$
\mathscr{J}_\mu=\mathscr{A}^\nu \times \mathscr{F}_{\nu\mu} + \mathscr{A}_\mu \times \mathscr{N} + i\partial_\mu \bar{\mathscr{C}} \times \mathscr{C} - i\bar{\mathscr{C}} \times D_\mu[\mathscr{A}] \mathscr{C}  +J_\mu 
$
with
$
 J_\mu^A  = -i \partial \mathscr{L}_{\rm matter}/\partial (\partial^{\mu} \psi_a)  (T_A \psi)_a  
$.
In fact, the total Lagrangian is invariant under color rotation, i.e., the global gauge transformation for $\Phi=\mathscr{A},\mathscr{C}, \bar{\mathscr{C}},\mathscr{N}$:
\begin{equation}
 \delta \Phi(x) := [\epsilon^C i Q^C, \Phi(x)] = \epsilon \times \Phi(x)  ,
\end{equation}
with the conserved Noether charge $Q^C := \int d^3x \mathscr{J}_0^C$ as the generator of the color rotation.

The original CF Lagrangian $\mathcal{L}_{\rm mYM}^{\rm tot}[\mathscr{A}_\mu,\mathscr{C},\bar{\mathscr{C}},\mathscr{N}]$ is written in terms of $\mathscr{A}_\mu, \mathscr{C}, \bar{\mathscr{C}}$ and $\mathscr{N}$.
The new theory is specified by $\mathcal{L}_{\rm mYM}^{\rm tot}[\mathscr{K}_\mu,\mathscr{C},\bar{\mathscr{C}},\mathscr{N}]$ written in terms of $\mathscr{K}_\mu, \mathscr{C}, \bar{\mathscr{C}}$ and $\mathscr{N}$ with the symmetry:
\begin{align}   
\begin{cases}
			{\boldsymbol \delta} \mathscr{K}_{\mu}(x) =  0
\\
			{\boldsymbol \delta} \mathscr{C}(x) 
=  -\frac{g}{2} \mathscr{C}(x) \times \mathscr{C}(x)  
\\
			{\boldsymbol \delta} \bar{\mathscr{C}}(x) =  i \mathscr{N}(x) 
\\
			{\boldsymbol \delta} \mathscr{N}(x) =  M^2 \mathscr{C}(x) 
\end{cases} 
 .
 \label{Sym}
\end{align}
Using the conserved Noether charge associated with the symmetry (\ref{Sym}), the physical state $\mathcal{V}_{\rm phys}$ is specified by the same form as the KO subsidiary condition \cite{KO78}
$Q_{\rm BRST}| {\rm phys} \rangle=0$ and/or $\bar{Q}_{\rm BRST}| {\rm phys} \rangle=0$.
Then the one-particle state $| \mathscr{K}_{\mu} \rangle= \mathscr{K}_{\mu} | 0 \rangle $ is a physical state, $| \mathscr{K}_{\mu} \rangle \in \mathcal{V}_{\rm phys}$, since $Q_{\rm BRST} | \mathscr{K}_{\mu} \rangle=[Q_{\rm BRST}, \mathscr{K}_{\mu}]| 0 \rangle  = -i{\boldsymbol \delta} \mathscr{K}_{\mu}| 0 \rangle =0$.

It was claimed in \cite{Ojima82} that the positive semidefiniteness of the physical subspace defined by this subsidiary condition  is violated in the massive non-Abelian gauge theories such as the CF model, as a consequence of which the physical $S$-matrix unitarity is invalidated.  
There a counter example was given in a quite simple and non-perturbative way,
a state $\Phi$ satisfying the KO subsidiary condition has the negative norm.
The argument is based on the asymptotic fields of the Heisenberg fields and the asymptotic (anti)BRST transformation which is linear in the asymptotic fields.
We do not use the asymptotic fields at all, since the existence of the asymptotic fields in QCD is problematic. 
Therefore, the state $\Phi$ given in \cite{Ojima82} does not satisfy the KO subsidiary condition imposed by the  BRST charge constructed from the Heisenberg fields, and it does not play the role of the counter example.

The violation of nilpotency of the modified (anti)BRST transformation is not the origin of nonunitarity against the preceding claims. 
Rather, violation of nilpotency is necessary for the massive vector theory to be unitary. Otherwise, the quartet mechanism \cite{KO78} works to eliminate both the scalar and longitudinal modes together with the ghost and antighost from the physical spectrum and there remain only two transverse modes as in the massless case, which is not anticipated in the massive case.  
In the massive case, only the scalar mode must be eliminated together with the ghost and anighost modes, leaving three physical massive modes, which is possible only when the nilpotency is violated. This is suggested from the relation:
\begin{align}
  {\boldsymbol \delta} {\boldsymbol \delta}  {\boldsymbol \delta} \bar{\mathscr{C}}(x) 
=   i {\boldsymbol \delta} {\boldsymbol \delta}  \mathscr{N}(x)
=  i M^2 {\boldsymbol \delta} \mathscr{C}(x) 
\not= 0
 . 
\end{align}
Note that $\mathscr{C}$, $\bar{\mathscr{C}}$ and $\mathscr{N}$ have the same mass $\sqrt{\beta}M$ and decouple in the (unitary gauge) limit $\beta \rightarrow \infty$ and only $\mathscr{K}_\mu$ remain as the massive spin 1 particle in the spectrum.

Moreover, this massive model has a smooth massless limit solving the mismatch in physical degrees of freedom between the massless vector (2) and  massive vector (3) cases.  
This issue already exists in the Abelian case and was solved by Nakanishi \cite{Nakanishi72}. 
Our model is regarded as a non-Abelian extension of the Nakanishi model. 
When $M \not= 0$, $\mathscr{K}_\mu |0 \rangle$ gives three independent physical states with positive norm, and $\mathscr{N} |0 \rangle$ becomes the unphysical state with negative norm.
When $M=0$, the scalar mode $\mathscr{N} |0 \rangle$ is a physical state, but it is unobservable due to the zero norm, while the longitudinal mode is unphysical.  Thus, the fact that observable independent degrees of freedom are 3 for $M \not= 0$ and 2 for $M=0$ is expressed by keeping the continuity with respect to $M$.

The eq.(\ref{K}) gives a transformation from $\mathscr{A}_\mu, \mathscr{C}, \bar{\mathscr{C}}$ and $\mathscr{N}$ to $\mathscr{K}_\mu$.
In order  to  obtain the new theory explicitly, we need the inverse transformation of eq.(\ref{K}).
But the inverse transformation can not be given in a closed form, since  (\ref{K}) is a nonlinear transformation. 
To perform the perturbative calculation, however, it is sufficient to know the order by order expression of the inverse transformation, which is obtained by the iterative procedure:
\begin{align}
 \mathscr{A}_\mu =&  \mathscr{K}_\mu + M^{-2} \partial_\mu \mathscr{N} 
+ gM^{-2} \mathscr{A}_\mu \times \mathscr{N} 
\nonumber\\
&- gM^{-2}  \partial_\mu \mathscr{C} \times i\bar{\mathscr{C}} 
- g^2 M^{-2} (\mathscr{A}_\mu \times \mathscr{C}) \times i \bar{\mathscr{C}} 
\nonumber\\
=& \mathscr{K}_\mu + M^{-2} \partial_\mu \mathscr{N} 
- gM^{-2} i\bar{\mathscr{C}} \times  \partial_\mu \mathscr{C}  
\nonumber\\
&+ gM^{-2} \mathscr{K}_\mu \times \mathscr{N} 
+ gM^{-4} \partial_\mu \mathscr{N} \times \mathscr{N} + O(g^2) .
\label{K2}
\end{align}
The propagator for $\mathscr{K}$ is obtained from the order $g^0$ terms, as in the Abelian case, and hence the propagator is not modified from the Proca case. 
However, the vertex functions among $\mathscr{K}_\mu, \mathscr{C}, \bar{\mathscr{C}}$ and $\mathscr{N}$ are modified from those of  $\mathscr{A}_\mu, \mathscr{C}, \bar{\mathscr{C}}$ and $\mathscr{N}$ in the original theory.
This fact has been overlooked in all the preceding studies.  
The failure of the physical unitarity reported in the preceding works \cite{DV70,SF70,Boulware70,CF76,CF76b,Ojima82,BSNW96} is based on the observation that   the vertices in the massive case are the same as the massless case. 
However, this observation is correct only if the relationship between the original field $\mathscr{A}_\mu$ to the massive field $\mathscr{K}_\mu$ is linear as in the Abelian case.  
This is not the case in the non-Abelian case, as our construction of the massive field $\mathscr{K}_\mu$ clearly shows. 
The vertices in the massive theory is modified in terms of the vector field $\mathscr{K}_\mu$ in addition to the FP ghost, antighost and the NL field, so that the physical unitarity holds even in the massive theory. 
This is the main reason why the physical unitarity was seemed to be violated so far. 

The original Lagrangian (\ref{mYM1}) is rewritten order by order of the coupling constant $g$ into
\begin{align}
\mathscr{L}_{\rm mYM}^{\rm tot} =& \mathscr{L}_0 + \mathscr{L}_1 + O(g^2) ,
\nonumber\\
\mathscr{L}_0 
=& - \frac14 (\partial_\mu \mathscr{K}_\nu - \partial_\nu \mathscr{K}_\mu)^2 + \frac12 M^2 \mathscr{K}_\mu \cdot \mathscr{K}^\mu 
\nonumber\\
&- \frac12 M^{-2} \partial^\mu \mathscr{N} \cdot \partial_\mu \mathscr{N}+ \frac12 \beta \mathscr{N} \cdot \mathscr{N} 
\nonumber\\
&- i\partial^\mu \bar{\mathscr{C}} \cdot \partial_\mu \mathscr{C} + \beta M^2 i\bar{\mathscr{C}} \cdot \mathscr{C} ,
\nonumber\\
\mathscr{L}_1 
=& - \frac{g}{2}  (\partial_\mu \mathscr{K}_\nu - \partial_\nu \mathscr{K}_\mu) \cdot (\mathscr{K}^\mu \times \mathscr{K}^\nu) 
\nonumber\\
&+ \frac{g}{2} M^{-4} (\partial_\mu \mathscr{K}_\nu - \partial_\nu \mathscr{K}_\mu) \cdot (\partial^\mu \mathscr{N} \times \partial^\nu \mathscr{N})
\nonumber\\
&- g M^{-2}  \mathscr{K}_\mu \cdot ( \mathscr{N} \times \partial^\mu \mathscr{N})
\nonumber\\
&+ \frac{g}{2} M^{-2} (\partial_\mu \mathscr{K}_\nu - \partial_\nu \mathscr{K}_\mu) 
\cdot (i\partial^\mu \bar{\mathscr{C}} \times \partial^\nu \mathscr{C} 
\nonumber\\& \quad\quad\quad\quad\quad\quad\quad\quad\quad\quad\quad\quad
- i \partial^\nu \bar{\mathscr{C}} \times \partial^\mu \mathscr{C}) 
\nonumber\\
&+ g   \mathscr{K}_\mu \cdot ( i\partial^\mu \bar{\mathscr{C}} \times   \mathscr{C} - i \bar{\mathscr{C}} \times \partial^\mu \mathscr{C}) 
\nonumber\\
&- gM^{-2} \partial_\mu \mathscr{N} \cdot ( i\partial^\mu  \bar{\mathscr{C}} \times \mathscr{C}  ) 
- g \beta  \mathscr{N} \cdot ( i\bar{\mathscr{C}} \times \mathscr{C}  ) ,
\end{align}
Then the Feynman rules for the new fields are given up to $O(g)$ according to the standard method. 
\\
(a) massive vector propagator
$
 \langle \mathscr{K}_\mu^A(-k) \mathscr{K}_\nu^B(k) \rangle 
$
\begin{equation}
   \frac{-1}{k^2-M^2+i\varepsilon} \left( g_{\mu\nu} - \frac{k_\mu k_\nu}{M^2} \right) \delta_{AB} , 
  \label{P1'}
\end{equation}
\\
(b) FP ghost propagator
$
 \langle \mathscr{C}^A(-k) \bar{\mathscr{C}}^B(k) \rangle 
$
\begin{equation}
   \frac{-1}{k^2-\beta M^2+i\varepsilon}   \delta_{AB} , 
  \label{P2'}
\end{equation}
\\
(c) NL field propagator
$
 \langle \mathscr{N}^A(-k) \mathscr{N}^B(k) \rangle 
$
\begin{equation}
   \frac{-M^2}{k^2-\beta M^2+i\varepsilon}   \delta_{AB} , 
  \label{P3'}
\end{equation}
\\
(d) $\mathscr{K}\mathscr{K}\mathscr{K}$ vertex 
$
\langle \mathscr{K}_\mu^A(k_1) \mathscr{K}_\nu^B(k_2) \mathscr{K}_\rho^C(k_3) \rangle
$
\begin{equation}
  ig f_{ABC} V_{\mu\nu\rho}(k_1,k_2,k_3) ,
  \label{V1'}
\end{equation}
where 
\begin{align}
  V_{\mu\nu\rho}(k_1,k_2,k_3)
:=& (k_1-k_2)_\rho g_{\mu\nu}+(k_2-k_3)_\mu g_{\nu\rho}
\nonumber\\&
+(k_3-k_1)_\nu g_{\mu\rho}  , 
\end{align}
\\
(e) $\mathscr{K}\bar{\mathscr{C}}\mathscr{C}$ vertex 
$
\langle \bar{\mathscr{C}}^A(k_1) \mathscr{C}^B(k_2) \mathscr{K}_\rho^C(k_3) \rangle
$
\begin{equation}
  ig f_{ABC} [M^{-2}(k_1^\rho k_2 \cdot k_3 -k_2^\rho k_1 \cdot k_3) + k_1^\rho-k_2^\rho], 
  \label{V2'}
\end{equation}
\\
(f) $\mathscr{K}\mathscr{N}\mathscr{N}$ vertex 
$
\langle \mathscr{N}^A(k_1) \mathscr{N}^B(k_2) \mathscr{K}_\rho^C(k_3) \rangle
$
\begin{equation}
  igM^{-2}f_{ABC} [M^{-2}(k_1^\rho k_2 \cdot k_3 -k_2^\rho k_1 \cdot k_3) + k_1^\rho-k_2^\rho], 
  \label{V3'}
\end{equation}
\\
(g) $\bar{\mathscr{C}}\mathscr{C}\mathscr{N}$ vertex 
$\langle \bar{\mathscr{C}}^A(k_1) \mathscr{C}^B(k_2) \mathscr{N}^C(k_3) \rangle$
\begin{equation}
  ig f_{ABC} [M^{-2} k_1 \cdot k_3 - \beta] . 
  \label{V4'}
\end{equation}

These rules should be compared with those for the original fields.
\\
(a)  vector propagator
$
 \langle \mathscr{A}_\mu^A(-k) \mathscr{A}_\nu^B(k) \rangle 
$
\begin{align}
& \frac{-1}{k^2-M^2+i\varepsilon} \left( g_{\mu\nu} - \frac{k_\mu k_\nu}{k^2+i\varepsilon} \right) \delta_{AB} 
\nonumber\\&
+ \frac{-\beta}{k^2-\beta M^2+i\varepsilon} \frac{k_\mu k_\nu}{k^2+i\varepsilon}   \delta_{AB} , 
  \label{P1}
\end{align}
\\
(b) FP ghost propagator
$
\langle \mathscr{C}^A(-k) \bar{\mathscr{C}}^B(k) \rangle 
= (\ref{P2'})
$
\\
(c) NL field propagator
$
 \langle \mathscr{N}^A(-k) \mathscr{N}^B(k) \rangle 
= (\ref{P3'})
$
\\
(c$^\prime$) the cross propagator
$
 \langle \mathscr{A}_\mu^A(-k) \mathscr{N}^B(k) \rangle 
$
\begin{equation}
   \frac{ik_\mu}{k^2-\beta M^2+i\varepsilon}   \delta_{AB} , 
  \label{P4}
\end{equation}
\\
(d) $\mathscr{A}\mathscr{A}\mathscr{A}$ vertex 
$
\langle \mathscr{A}_\mu^A(k_1) \mathscr{A}_\nu^B(k_2) \mathscr{A}_\rho^C(k_3) \rangle
= (\ref{V1'})
$
\\
(e) $\mathscr{A}\bar{\mathscr{C}}\mathscr{C}$ vertex 
$
\langle \bar{\mathscr{C}}^A(k_1) \mathscr{C}^B(k_2) \mathscr{A}_\rho^C(k_3) \rangle
$
\begin{equation}
  ig f_{ABC} k_1^\rho . 
  \label{V2}
\end{equation}

\begin{figure}[t]
\begin{center}
\includegraphics[height=4.5cm,width=6.5cm]{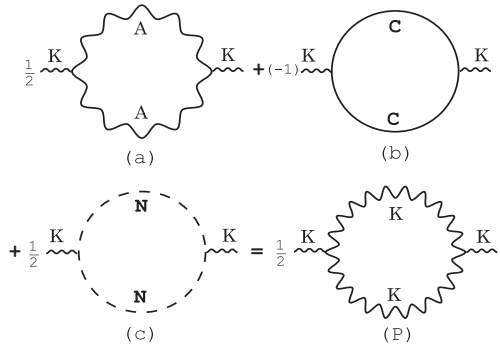}
\end{center}
\vspace{-0.3cm}
\caption{ 
Diagrams for checking the physical unitarity, (a) original gluon, (b) ghost and antighost, (c) NL field, (P) physical gluon
}
\label{fig:diagram}
\end{figure}

We proceed to demonstrate the physical unitarity for the simplest one loop case of Fig.~1 examined in \cite{DV70}. 
We chose $\beta=1$ for simplicity.
The imaginary part of the diagram (a) is given from (\ref{P1}) and   (\ref{V1'}) by
\begin{align}
(a)=  \frac12 igf_{ABC} V_{\mu\nu\rho}(k_1,k_2,k_3) igf_{ABC^\prime} V_{\mu^\prime\nu^\prime\rho^\prime}(k_1,k_2,k_3)
\nonumber\\
\times \pi \delta(k_1^2-M^2) g^{\mu\mu^\prime} \pi \delta(k_2^2-M^2) g^{\nu\nu^\prime} \epsilon^{(j)\rho}(k_3) \epsilon^{(k)\rho^\prime}(-k_3),
\end{align}
where $\epsilon^{(j)}(k)$ ($j=1,2,3$) denotes one of the three polarization vector of the massive vector field.
The physical unitary requires the imaginary part of the second order diagram to be equal to the physical part $(a)^{P}$ which is obtained from $(a)$ by the replacement:
\begin{align}
g^{\mu\mu^\prime}  &\rightarrow    g^{\mu\mu^\prime} - \frac{k_1^\mu k_1^{\mu^\prime}}{M^2}    
 = - \sum_{j=1}^{3} \epsilon^{(j)\mu}(k_1) \epsilon^{(j)\mu^\prime}(k_1)  ,
 \nonumber\\
g^{\nu\nu^\prime}  &\rightarrow    g^{\nu\nu^\prime} - \frac{k_2^\nu k_2^{\nu^\prime}}{M^2}    
 = - \sum_{k=1}^{3} \epsilon^{(k)\nu}(k_2) \epsilon^{(k)\nu^\prime}(k_2)  .
\end{align}
The difference $(a)-(a)^{P}$ between $(a)$ and $(a)^{P}$ must be cancelled by the imaginary parts of the other diagrams in the second order, coming from unphysical fields, i.e., ghost, antighost and NL field. 
First we consider the imaginary part of the diagram (b) with a closed loop of the FP ghost and antighost.
By using the Feynman rules given above, and noting that $k_1^2=k_2^2=k_3^2$ and $k_{1\mu} \epsilon^{(j)\mu}(k_1)=k_{2\mu} \epsilon^{(j)\mu}(k_2)=k_{3\mu} \epsilon^{(j)\mu}(k_3)=0$, we find that the ghost contribution is precisely of the same form as the  difference $(a)-(a)^{P}$ and comes with the opposite sign.  
Consequently, the difference $(a)-(a)^{P}$ cancels against half the ghost contribution:
$(a)-(a)^{P}=-\frac12 (b)$,  as found in \cite{DV70} and emphasized in \cite{DTT88}. 
Hence, the ghost loop gives the excessive contribution $\frac12 (b)$ violating the physical unitarity:
$(a)+(b)=(a)^{P}+\frac12 (b) \not= (a)^{P}$.

This situation is remedied by including the imaginary part of the diagram (c) with a closed loop of the NL field. 
By taking into account the fact that the vertex (e) $\mathscr{K}\bar{\mathscr{C}}\mathscr{C}$ is the same as the vertex 
(f) $\mathscr{K}\mathscr{N}\mathscr{N}$ up to the factor $M^2$,  
it is easy to find the relation $(c)=-\frac12 (b)$, which leads to $(b)+(c)=\frac12 (b)$.
Hence, the cancellation occurs, $[(a)-(a)^{P}]+[(b)+(c)]=-\frac12 (b)+\frac12 (b)=0$.
Thus we have shown the physical unitarity,  $(a)+(b)+(c)=(a)^{P}$. 
The 1/2 factor is essential for unitarity in the massive vector theory. 
In the proposed model, the NL field plays the same role as the St\"uckelberg field which cancels half the ghost contribution to ensure physical unitarity.

The proposed model opens a path of resolving the long-standing problem of reconciling physical unitarity with renormalizability without Higgs fields,
and it can be useful as a regularizations scheme for infrared divergences in non-Abelian gauge theories.

{\it Acknowledgements}\ ---
This work is  supported by Grant-in-Aid for Scientific Research (C) 21540256 from Japan Society for the Promotion of Science (JSPS).



\begin{thebibliography}{99}
\bibitem{Higgs66}
P.W. Higgs,
Phys. Rev. {\bf 145}, 1156
 (1966).


\bibitem{tHooft71}
G. 't Hooft,
Nucl. Phys. B{\bf 35}, 167
 (1971).


\bibitem{YM54}
C.N. Yang and R.L. Mills,
Phys. Rev. {\bf 96}, 191
 (1954).


\bibitem{DV70}
H. van Dam and M.J.G. Veltman,
Nucl. Phys. B{\bf 22}, 397
 (1970) 


\bibitem{SF70}
A.A. Slavnov and L.D. Faddeev, 
Theor. Math. Phys. {\bf 3},  312
 (1970) [Teor. Mat. Fiz. {\bf 3}, 18
 (1970)]. 
\\
A.A. Slavnov,
Theor. Math. Phys. {\bf 10},  201
 (1972) [Teor. Mat. Fiz. {\bf 10}, 305
 (1972)]. 


\bibitem{Boulware70}
D.G. Boulware, 
Annals Phys. {\bf 56}, 140
 (1970). 


\bibitem{CF76}
G. Curci and R. Ferrari,
Nuovo Cim. A{\bf 32}, 151
 (1976).


\bibitem{CF76b}
G. Curci and R. Ferrari, 
Nuovo Cim. A{\bf 35}, 1
 (1976), Erratum-ibid. A{\bf 47}, 555 (1978). 


\bibitem{Ojima82}
I. Ojima, 
Z. Phys. C{\bf 13}, 173
 (1982).


\bibitem{BSNW96}
J. de Boer, K. Skenderis, P. van Nieuwenhuizen and A. Waldron,
Phys. Lett. B{\bf 367},  175
 (1996).
 

\bibitem{FMTY81}
T. Kunimasa and T. Goto,  
Prog. Theor. Phys. {\bf 37}, 452
 (1967). 
\\
T. Fukuda, M. Monda, M. Takeda and Kan-ichi Yokoyama,
Prog. Theor. Phys. {\bf 66}, 1827
 (1981).


\bibitem{DTT88}
R. Delbourgo, S. Twisk and G. Thompson,
Int. J. Mod. Phys. A{\bf 3}, 435
 (1998).

\bibitem{RRA04}
H. Ruegg and M. Ruiz-Altaba,
Int. J. Mod. Phys. A{\bf 19},  3265
 (2004).


\bibitem{Cornwall82}
J.M. Cornwall, 
Phys. Rev. D{\bf 26}, 1453
 (1982). 
\\
J.M. Cornwall and A. Soni,  
Phys. Lett. B{\bf 120}, 431
 (1983). 


\bibitem{decoupling}
 Ph. Boucaud, J.P. Leroy, A. Le Yaouanc, J. Micheli, O. Pene and J. Rodriguez-Quintero, 
JHEP 0806, 099 (2008).
\\
 A.C. Aguilar, D. Binosi and J. Papavassiliou,
Phys. Rev. D{\bf 78}, 025010 (2008).


\bibitem{Baulieu85}
L. Baulieu,  
Phys. Rept. {\bf 129}, 1
 (1985).


\bibitem{KO78}
T. Kugo and I. Ojima,
Phys. Lett. B{\bf 73}, 459
 (1978).

 
\bibitem{Nakanishi72}
N. Nakanishi,
Phys. Rev. D{\bf 5},  1324
 (1972).

 






\end{thebibliography}
\end{document}